\documentclass[10pt]{bmc_article}

\usepackage{cite} 
\usepackage{url}  
\usepackage{ifthen}  
\usepackage{multicol}   
\usepackage[utf8]{inputenc} 
\usepackage{graphicx}
\usepackage{booktabs}
\usepackage{lscape}
\usepackage{fancyhdr}
\usepackage{blindtext,tikz}
\usetikzlibrary{calc}
\usepackage{graphicx,eso-pic,xcolor}

\newboolean{publ}

\newenvironment{bmcformat}{\baselineskip15pt\sloppy\setboolean{publ}{false}}{\baselineskip15pt\sloppy}

\begin{document}

\begin{bmcformat}

\title{Rigidity and flexibility in protein-protein interaction networks: a case study on neuromuscular disorders}


\author{Ankush Sharma\correspondingauthor$^{1}$
        \email{Ankush Sharma \correspondingauthor- ankush.sharma@na.icar.cnr.it}, Maria Brigida Ferraro$^{2}$%
        \email{Maria Brigida Ferraro - mariabrigida.ferraro@uniroma1.it}, Francesco Maiorano$^{1}$%
        \email{francesco.maiorano@na.icar.cnr.it}, 
        Francesca Del Vecchio Blanco$^{3}$
        \email{Francesca Del Vecchio Blanco - francesca.delvecchioblanco@unina2.it} and Mario Rosario Guarracino$^{1}$%
         \email{Mario Rosario Guarracino - mario.guarracino@cnr.it}
         }

\address{$^{1}$Laboratory for Genomics,Transcriptomics and Proteomics, High Performance Computing and Networking Institute,\\ \hspace{4pt}National Research Council, Via P. Castellino, 111, Naples, Italy\\
$^{2}$Department of Statistical Sciences, Sapienza University of Rome, P.le A. Moro 5, Rome, Italy\\
$^{3}$Department of Biochemistry, Biophysics and General Pathology,Second University of Naples, Italy.}

\maketitle

 \clearpage
\begin{abstract}

\subsection*{Background}
Mutations in proteins can have deleterious effects on a protein's stability and function, which ultimately causes particular diseases. Genetically inherited muscular dystrophies include several genetic diseases, which cause increasing weakness in muscles and disability to perform muscular functions progressively. Different neuro-muscular diseases are caused by different types of mutations in the gene coding. Mutations in genes make defunct proteins or none at all.  Defunct or missing protein interactions in human proteome may cause a stress to its neighboring proteins and subsequently to modules it is involved in. Network biology is utilized to gain knowledgeable insights on system properties of complex protein-protein interaction maps governing affected cellular machinery due to disease causing mutations. We therefore aimed to understand the effects of mutated proteins on interacting partners in different muscular dystrophies.\\
\subsection*{Results}
We investigated rigidity and flexibility of protein-protein interaction subnetworks associated with causative mutated genes showing high mean interference values in muscular dystrophy. Rigid component related to Eukaryotic Translation Elongation Factor 1 Alpha 1 (EEF1A1) subnetwork and members of 14.3.3 protein family formed the core of network showed involvement in molecular function related to protein domain specific binding. Core nodes of core modules showed high modular overlapping and bridgeness values.  The subnetworks showing highest flexibility comprised of seed nodes Calcium channel, voltage-dependent, L type, alpha 1S subunit (CACNA1S) and calmodulin 1 (CALM1) showing functionality related to Voltage-dependent calcium channel. The interconnected subnet of proteins corresponding to known causative genes having large genetic variants are shared in different  Muscular dystrophies (MDs) inferred towards comorbidity in diseases.\\ 
\subsection*{Conclusion}
 The studies demonstrates core network of MDs as highly rigid component constituting of large intermodular edges and interconnected hub nodes suggesting high information transfer flow.The core skeleton of the network are organized in protein binding and  protein specific domain binding.This suggests neuro-muscular disorders may initiate due to interruption in molecular function related with the core and its aggression may depend on the tolerance level of the networks.

\end{abstract}
 

\ifthenelse{\boolean{publ}}{\begin{multicols}{2}}{}

\clearpage
\section*{Introduction}
Neuro-muscular diseases, such as muscular dystrophies (MDs), are associated with muscle weakness, muscle atrophy and a progressive cardiac dysfunction over time.  Any muscle can be affected by this condition, however it is most prominent in limb-girdle and proximal muscle groups with varied involvement of distal muscles \cite{AmiEtAl03}.
Myopathies in effect for a long period may produce loss of absolute volume of muscle and these conditions are associated with muscle wasting \cite{BonSan13}.
Genetically heterogeneous MDs range from severe to benign forms such as Ducchene to Limb girdle muscular dystrophy (mild). Mitochondrial abnormalities are associated with ocular myopathies, whereas metabolic disorders are involved in the acquired chronic inflammatory myopathies dermatomyositis and polymyositis \cite{AmiEtAl03}.

Most diseases are caused by mutations in more than one gene which can either be dominant or recessive.
In MDs, Such dominant and recessive pathological defects are shared by a subset of genes involved in diverse mechanisms related to muscle degeneration and weakness. Such mechanisms orchestrate many intricate and common biological pathways  \cite{McnPyt07}. 
Biological pathways consist of a set of dependent actions carrying out a specific function. They take place in cells among molecules such as proteins, metabolites and enzymes. A particular function can be hampered if one or more components of a biological pathway fail to perform. Damaging mutations in proteins can be considered as one of such cases,  aborting their functionality and related pathways. Missense mutations in muscular dystrophies are often associated with neuro-muscular abnormalities and cognitive impairment \cite{HarEtAl11,SinEtAl10}. 

	\subsection*{Biological networks }
Multiple biological pathways lack boundaries,  often are interconnected, and work together to accomplish tasks. The interconnected component of biological pathways is called a biological network. Networks are valuable prototypes for analyzing the complexity in cellular environments and the interactions, which influence the normal functionality of the cells. 

Biological networks exhibit modular organization dependent on functions. Modules are sets of nodes that share many edges, and are loosely connected to the rest of the network, representing densely associated entities. Core nodes of a module have large number of edges to other proteins within modules and are essential to  its functions. Modules exhibit fuzzy boundaries and are interconnected to perform wide variety of functions in cells. Perturbed components in a network, such as in the case of mutated proteins, can influence the coherent overlapping modules in human proteome \cite{CseEtAl11}.
Proteins linking different modules together are important for inter-modular communications and show high bridgeness values \cite{ZhuEtal07}. Modules in protein-protein interaction networks of yeast partially disintegrate upon stress,  removing important inter modular edges, thus preventing flow of information \cite{MihCse11}. Moreover, if a module contains proteins with unknown function, the functional characterization may help in determining functional prediction of those proteins. \cite{NewGir03,RivGal03}. 

Better understanding of structural functional aspects can be studied taking dynamics of networks into account, along with network topology.  
The latter provides understanding of network architecture. Biological networks show non random degree distribution and small world property. Non random degree distribution also known as scale-free degree distribution, in which large number of nodes have fewer edges and few nodes having many edges are known as hub nodes. Their targeted deletion disrupts the network structure \cite{BarAlb99,JeoEtAl01}.

Hub proteins serve as common edges and mediate short path lengths between other edges. Shortest path length is a distance between two nodes and median of the means of the shortest path lengths connecting each nodes to all other vertices is known as characteristic path length. Small world networks, in which any two nodes in the networks can be connected with short paths, exhibit smaller diameter, small characteristic path length, and high clustering coefficient \cite{watstr98}. Clustering coefficient ranges from 0 to 1, and provides a measure of the degree to which nodes tend to cluster in network \cite{ZhuEtal07}. Betweenness centrality is the number of shortest paths from all vertices to all others that pass through that node. It measures traffic loads through one node  as information flows over a network primarily following the shortest available paths. High betweenness centrality proteins behaves as a bottlenecks in protein-protein interaction networks. Bottleneck proteins regulate most of the informational flow, hence indicates the essentiality of proteins \cite{Nar05,YuKIm07}. Clique is a set of nodes which have all possible ties among themselves. A maximal clique is a clique that is not contained in any other clique.

	\subsection*{Propagation, rigidity and flexibility in networks}
The diseased state can be characterized as a malformed propagation state among constituting modules. Dynamics of modules from rigid to flexible state depend upon the environmental cues and the internal structure of the networks. 
Removal of protein or protein complexes may significantly alter the network, affecting flow of information, efficiency and adaptability. Adaptability is  associated with flexibility of the networks, whereas rigidity is associated with the memory \cite{GasEtAl12}. Rigidity and flexibility can be assessed by combinatorial graph theory and it is described by degrees of freedom and number of over-constraints associated with the nodes in the network \cite{FoxStr12}.

	\subsection*{Our approach}
In this article, we focused on ({\em i}) assessing the snapshot of dynamics of network propagation and interference from pairs of hub and essential mutated proteins causative in different neuro-muscular disorders; ({\em ii}) evaluating rigidity and flexibility (tolerance level) in protein-protein interaction networks of the most interfered sub-networks in muscular dystrophies. ({\em iii}) determining the modular organization and, ({\em iv}) characterizing molecular function of modules.  

A systematic characterization of MDs has to take into account the overlap of different  traits and the heterogeneity of cellular processes  for many of the genes involved.  The analysis and classification of multiple genes and their mutations is still challenging in terms of cost and time, even with the advent of sequencing technologies \cite{BarRit11}. The proposed strategies overcomes the limits of existing solutions which compare protein-protein interaction networks of disease and control states, solely on topological characteristics. We are able to predict which sub-networks are more rigid in presence of such mutations. To best of our knowledge, we investigated for the first time, the rigidity and flexibility issues in sub-networks of human proteome.

	\section*{Methods}

The neuro-muscular diseases are broadly classified in 13 groups based on previous studies \cite{Kap11}. We construct the protein interaction map of causative genes involved in the disease using a dataset by Center for Biomedical Computing at University of Verona \cite{ScaEtAl12}. The dataset is extracted from various databases storing high throughput methods and experimentally known interactions \cite{ChaEtAl07, KesEtAl09, SchEtAl09, StaEtAl06}, and it is manually curated and updated.  Information on disease causing variants is downloaded from Leiden Open Variant Database \cite{FokEtAl11}. Networks are visualized using Cytoscape and Gephi  \cite{SmoEtAl11, BasEtAl09}. Network layout is performed using force directed graph drawing \cite{FruEtAl91}. Centrality measure of nodes, which gives relative importance of nodes within a network, are calculated using Netanalyzer \cite{JeoEtAl01, AssEtAl08} and in-house R and python scripts. 
Hub and key proteins of the interconnected seed network are computed using degree distribution, betweenness centrality (BC), maximal clique centrality (MCC), and bottleneck nodes \cite{LinEtAl08}. 

	\subsection*{Network propagation}
The interference is the measure of overlapping flow among nodes \cite{StoYu09}. It is obtained visiting each node by random walks initiating at different sources. Larger interference implies wider overlap between flows  originating from different sources and small interference signifies little overlap. The emitting model is obtained using {\tt qmbpmn-tools}  \cite{StoEtal12}, which is initiated with every possible pair of proteins  in the first order network, to compute mean interference and maximum visits.

	\subsection*{Module detection and functional characterization}

ModuLand framework analyzes the overlapping modules in networks using bridgeness and overlapping values. It uses Proportion Hill module membership assignment method and NodeLand influence function algorithm \cite{SzaEtAl12,KovEtAl10}. In ModuLand framework, modularity of a network is computed by determining community centrality by summing up influence zones containing a given edge. Overlapping modules are identified on the basis of hills on community centrality landscape, and each node of the network is assigned to the module with different strength \cite{SzaEtAl12,KovEtAl10}. Overlap values for nodes demonstrates the effective number of modules to which they are assigned, and bridgeness values are high, if nodes show larger overlap between many module pairs. Modules are named after the core node of the module. BiNGO plug-in is used for functional characterization of the modules \cite{MaeEtAl05}. Molecular function is assigned on consensus basis to the modules based on p-values and involvement on core nodes from top ten core nodes in the molecular function.

\subsection*{Rigidity and flexibility}
Rigidity and flexibility issues were previously studies at a structural level of proteins through various methodologies \cite{Uversky13,CosEtAl13, ShaEtAl12, OldEtAl08}. We investigated the properties characterizing different states such as rigidity and flexibility of a complex system upon perturbation. Rigidity and flexibilty for subnetworks showing maximum interference values and first order network are analyzed using pebble game algorithm in KINARI-lib \cite{FoxEtal11,FoxStr12}.
The pebble game algorithm computes total number of degrees of freedom and overconstrained regions in a graph. This algorithm requires sparsity parameters $(k,l)$ on protein-protein interaction networks to be analyzed. Pebble game algorithm on 2D bar joint framework is guaranteed for rigidity to all $(k,l)$-sparse graphs for $k$ and $l$ such that $l \in (0,2k)$. Initially, $k$ pebbles are posed on each vertex with no edges and then one of pebbles is displaced from $i$ after adding an edge $ij$ towards $j$, if at least $l+1$ pebbles are between vertices $l$ and $j$. In continuation to this, $ij$ is reversed and pebble is moved from $j$ to $l$, if pebble is on $j$ and $ij$ edge exists in pebble game's graph. Any subset of $n'$ vertices spans at most $kn'-l$ vertices in $(k,l)$ graph is called sparse whereas it is called as tight or rigid if it has $n$ vertices and $kn-l$ edges \cite{Jac97,FelEtAl08,JacTho95}.
We compute the index (overconstraints - degree of freedom) divided by average degree of subnetwork to avoid size bias of the network in calculation of  DOF's and over-constrained regions. High positive values obtained for the subnetwork are associated to high rigidity, and negative values with flexibility. This is based on the fact that higher degrees of freedom represent higher flexibility while higher over-constrained regions or nodes indicate rigidity of the network.

\section*{Results}

	\subsection*{Network Topology}
There are 206 proteins affected by causative mutations in genes responsible for MDs, which are experimentally known to show protein-protein interaction. Such 
proteins form an interconnected component (seed network) of the protein-protein interaction map composed of  137 nodes and 307 edges. The first order network obtained from the 206 proteins is composed of 4076 proteins and 133847 edges. Both networks satisfy scale free property, following the power law in degree distribution \cite{BarAlb99}. Scale free property infer the robustness of such networks against random failures of the nodes (Figure \ref{fig:1}). The seed network has a clustering coefficient of 0.28 whereas first order network forms clusters with clustering coefficient 0.33. The small characteristic path length of 3.95 and 2.53 for interconnected seed network and first order network respectively shows average efficiency of transmission of information in network in less than 4 steps. Essential and hub nodes in the interconnected component is calculated using centrality statistics, which gives measure of load and linkedness of the nodes [Table \ref{tab:1}]. Large number of genetic variants are associated with hub proteins and essential nodes {(Supplementary Table S1)}  

\begin{figure}[!ht]
	\centering
	\includegraphics[scale=0.9]{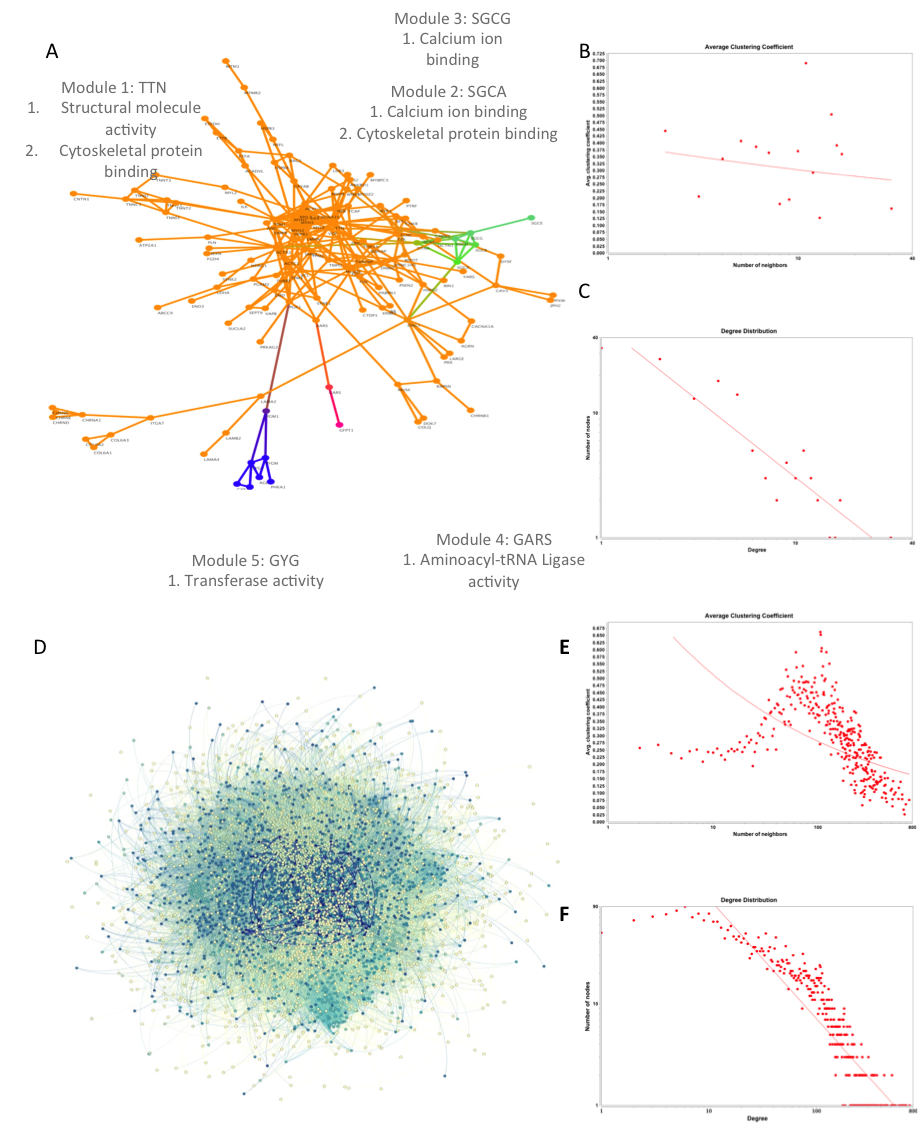}
\caption{\textit{ (A) Modular organization of giant component of protein-protein interaction network of seed mutated nodes in different muscular dystrophies. (B) Average clustering Coefficient distribution of nodes (C) Degree distribution of proteins in networks. (D) First order network of 206 proteins showing causative mutations (E) Average clustering coefficient  distribution of  proteins (F) Degree distrubution of proteins in first order network.}  }
	\label{fig:1}
\end{figure}

\begin{table}[!ht]
\centering

   \renewcommand{\arraystretch}{1.3} 
\caption{Key and Hub nodes computed based on centrality statistics: Degree, Betweenness Centrality, Bottleneck and Maximal Clique Centrality. The key nodes in bold are considered on consensus basis if occurred at least twice in detection.}
 \vspace{0.6cm}
\begin{tabular}{cccc}
        
      \hline
     
Degree & Betweenness  & Bottleneck & Maximal Clique \\
	    &	Centrality	    &  		       &Centrality	  \\
  \hline
   
 \textbf{TTN} 		& \textbf{DAG1}     		& \textbf{ TTN}        		& \textbf{TTN}	       \\
 \textbf{TPM1}  		&\textbf{TTN}              	& \textbf{ DAG1}      		& \textbf{DES}         \\
 \textbf{LMNA}  	 	&\textbf{VCL}              	& \textbf{ACTA1}      		& \textbf{TPM1}          \\
 \textbf{ACTN2}  	& 		ACTA1               	& \textbf{TPM1}      	 	& \textbf{LMNA}        	\\
 \textbf{DES}   		&\textbf{LAMA2}                  & \textbf{HSPB1}      		& \textbf{MYLK2}          \\
      	ACTA1 		& 		PGK1                 & LAMA2      			& \textbf{ACTN2}         \\
 \textbf{FLNA}  		&\textbf{FLNA}               	& \textbf{FLNA}      		& MYL3                    \\
 \textbf{HSPB1} 		&		 ITGA7               &  PGK1        			& MYH7                      \\
 \textbf{VCL}   		&		 DMD                 &  ITGA7      			& CACNA1S                   \\
 \textbf{MYLK2}  	& \textbf{TPM}		  	&\textbf{VCL}       		 & MYH2 \\
      \hline
\end{tabular}
 \label{tab:1}
\end{table}

\subsection*{Functional characterization} 
 The interconnected seed network is modularly organized around five overlapping modules. Hub node TTN formed core module of the seed network with molecular function related to structural molecule activity  ($p = 3.9E-7$, GO-id 5198), along with calmodulin binding ($p = 2.8E-5$, GO-id 5516). Module SGCA and Module SGCG are involved in calcium ion binding ($p=5.9E-4$ and $p=4.3E-4$, GO-id 5509), respectively, which is known to be involved in diseased state of Sarcoglycanopathies\cite{WhiEtAl05}. Module GYG1 is involved in catalytic activity such as transferase activity  ($p=2.2E-4$, GO-id 16740). Module GARS comprises of only 3 proteins with functionality related to ligase activity ($p=6.9E-3$, GO-id 16874). All those modules are depicted in {(Figure 1(A))}.
Nineteen overlapping modules characterize the first order network on consensus basis with function related to protein, DNA binding, transferase activity and structural molecule activity {(Table \ref{tab:2})}. The top ten core nodes governing the modular function is listed in {(Supplementary Table S2)}. 
The interconnected hub nodes constitutes the core skeleton of the first order network as well as central module EEF1A1.{(Supplementary Figure S1)}.

\begin{figure}[!ht]

	\centering
	\includegraphics[scale=.6]{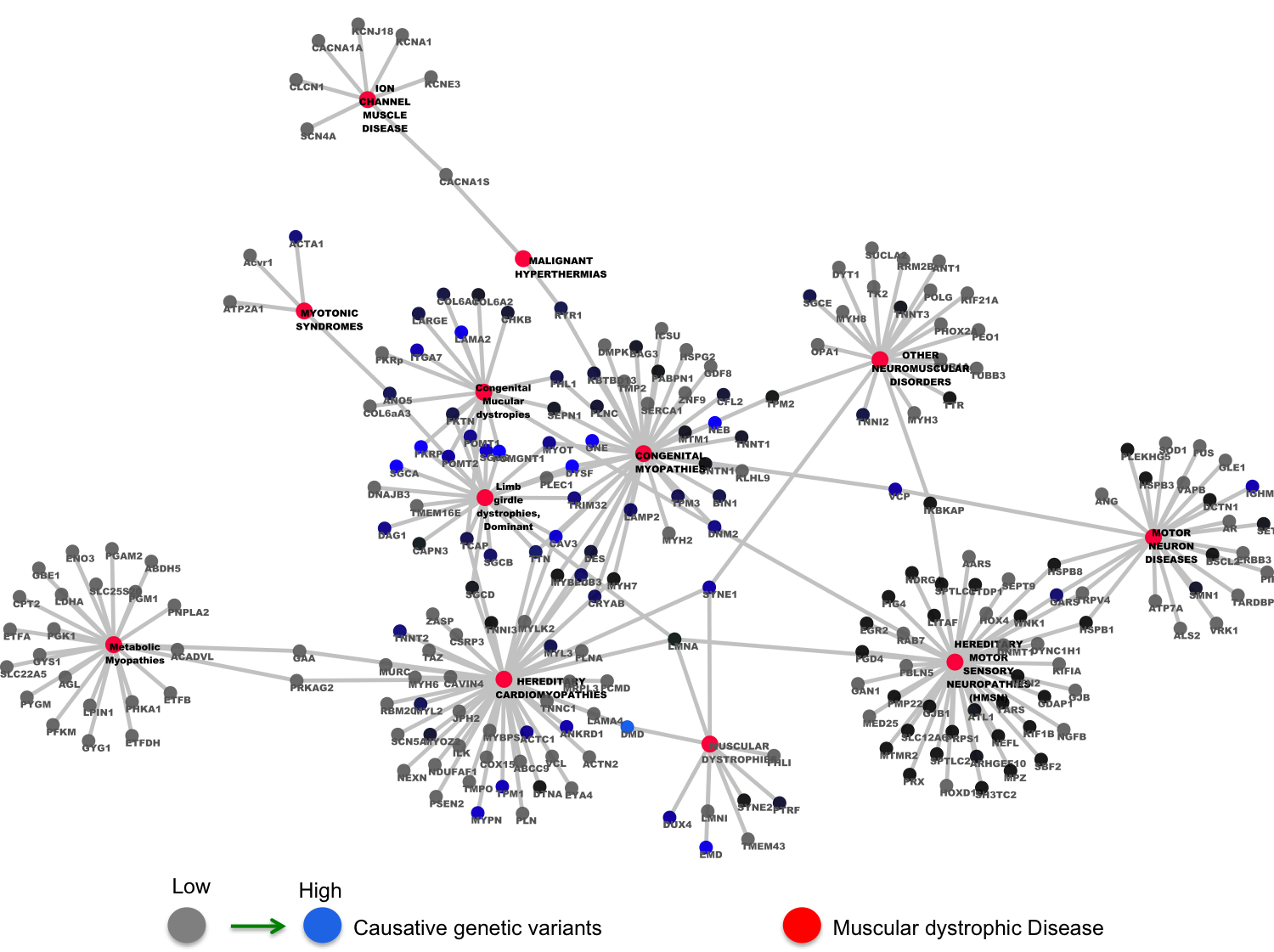}
\caption{\textit{Relationship between disease and proteins corresponding to causative mutated genes. Red nodes are diseases and Grey $->$ blue is low to high value of genetic variants.}}
	\label{fig:2}
\end{figure}
\clearpage
	\subsection*{Protein-disease association} 
The interaction map of formed by the seed mutated proteins and the thirteen neuro-muscular diseases is composed of 268 nodes with 285 edges. Disconnected components of the network relate to congenital myasthenic syndromes,linked with 13 proteins. The giant meta-network of the remaining twelve neuro-muscular disorders comprise 262 nodes and 270 edges. The network in {(Figure \ref{fig:2})} depicts proteins showing higher genetic variants shared by many different muscular diseases.
Congenital myopathies and limb girdle dystrophies, dominant (LGD, dominant) share numerous interacting partners with congenital myopathies and hereditary cardiomyopathies. LMNA protein shows involvement in four different muscular dystrophic diseases. The DMD protein connected with large number of interacting partners in first order protein interaction map, it has highest genetic variants, and it is specifically linked to muscular dystrophy and hereditary cardiomyopathies in protein-disease interaction map (Figure \ref{fig:2}). It is already very well known fact that cardiac disease is a clinical manifestation related to muscular dystrophies.\cite{HerEtAl10}.

	\subsection*{Network propagation from key pair of causative proteins in muscular dystrophies }
Modules exhibit fuzzy boundaries as discussed earlier and it is therefore hard to understand the rigidity and flexibility associated with them. In order to understand the rigidity and flexibility associated with the modules, we compute the interference (flow overlap) on first order network from pair of key proteins of seed network listed in {(Table \ref{tab:1})}. The network propagation initiating from these nodes is obtained with the using emitting model of {\tt qmbpmn-tools}, which calculates interference in the first order network {(Supplementary Table S3)}. From the top forty proteins showing maximum interference from each pair of mutated proteins. We then ranked them on quartiles of the computed mean interference. In {(Table \ref{tab:3})}, proteins with high bridgeness in the network ($>$1) are shown as receiving interference.

   Maximum number of visits producing highest mean interference value is observed in the core node YWHAZ, present in 3 different modules and responsible for molecular function related to protein domain specific binding ($p=4.0E-06$, GO-id 19904). The other proteins with mean interference greater than quartile percentage of 90 {[see Supplementary Table \ref{supptab:4}]} and large number of visits show diverse functionality ranging from localization ($p=4.6E-2$, GO-id 51179) to cellular component organization ($p=2.3E-2$, GO-id 16043) and intracellular transport ($p=4.6E-2$, GO-id 46907). CACNA1S receives largest mean interference value from the single duplet of DES/TPM1, which shows involvement in voltage gated calcium channel activity and skeletal muscle adaptation ($p=3.8E-2$, GO-id 43501). DES/TPM1 proteins are functional in structural constituent of cytoskeleton ($p=1.8E-3$, GO-id 5200).

	\subsection*{Rigidity and flexibility in protein protein interaction networks}
We focus on assessing network rigidity and flexibility issues and determine the tolerance level of the sub-networks of the proteins showing maximum interference. Extracted subnetworks show heterogeneity and varying clustering coefficient ranging from 0.531 to 0.994 with varied clustering pattern of proteins. {[Supplementary table \ref{supptab:5}]}. This provide an overview on the global network, and the possibility to determine which subnetwork is susceptible to affect the functionality of the modules. KINARI-lib computed The rigidity/flexibility for whole network for $k=2$ and $l=3$ showing 24509 degrees of freedom and 130751 over constraints. Variable rigidity/flexibility in subnetworks shed light on varied tolerance level in structural and functional integrity based on degrees of freedom and over constraints {[Figure \ref{fig:3}]}.
\begin{figure}[!ht]
\centering
\includegraphics[scale=.5]{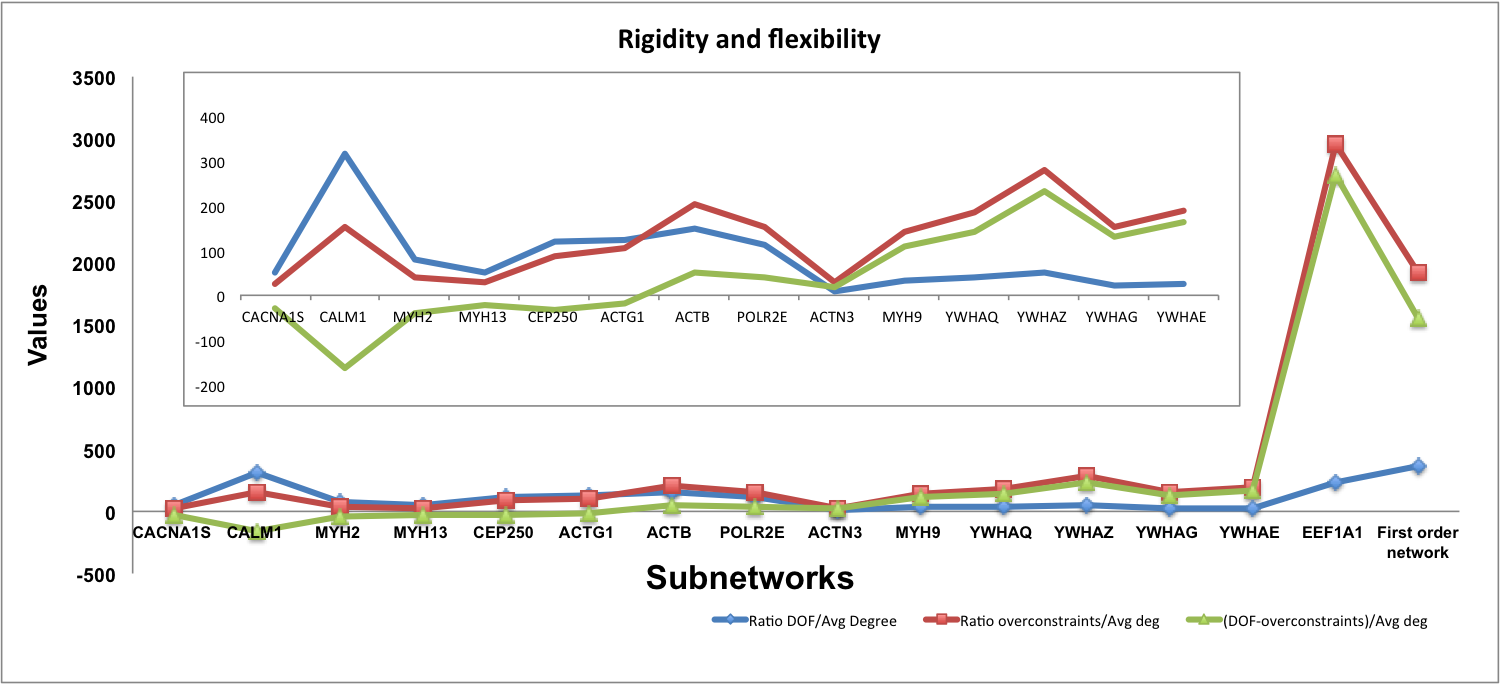}	
\caption{ \textit{Showing  ratio of DOF (Degree of Freedom) and average degree, overconstraints and average degree and difference of DOF and overconstraints with respect to average degree to overcome size bias of the network $k =2$ and $l=3$ for the subnetworks of nodes listed in Table 1 which showed mean interference values calculated from emitting model of ITMprobe.  Positive values in green line indicate rigidity and negative values indicate flexibility. Higher the positive value higher the network is rigid and vice versa. }}
 	\label{fig:3}
\end{figure}

Subnetwork of core node eukaryotic translational factor 1 alpha (EEF1A1) of core module EEF1A1 shows rigidity with many folds over-constrained nodes over degrees of freedom. This subnetwork has even higher constrained nodes as compared with first order network of the seed nodes. This module function is related to protein binding  {($p=2.6E-02$)}.  

The rigid sub networks corresponding to 14-3-3 proteins family, which binds to functionally diverse signaling proteins, are also rigid components of the network. The 14.3.3 family protein subnetworks (YWHAE, YWHAG, YWHAZ and YWHAQ) consitituting the core nodes of module Heat Shock 70kDa Protein 8 (HSPA8) in this network share many interacting partners. The rigidity results for over-contrained regions over degree of freedoms in subnetworks are consistent for all possible value analyzed, and for all the possible combination of $k=1,\dots,6$ and $l=1,\ldots,6$. {(Supplementary Figure S2)}.
The CACNA1S and CALM1 sub-networks demonstrate maximum flexibility with minimum over-constraints with nodes. With respect to degrees of freedom, they show involvement in functionality related voltage-gated calcium channel activity ($p=2.2E-02$, GO–id 5245) and calcium ion binding ($p=2.9E-3$, GO-id 5509). The flexibility in this subnetworks indicates functional flexibility of calcium, that is known to have various functions in our body and countering external influences for proper functioning in cellular environment calcium synergists are needed. 

\clearpage
	\section*{Discussion}
	
	The rigidity and flexibility issues in networks are much less studied in context of biological networks. Biological networks have modular organization \cite{BarZol04} without proper boundaries therefore, it is hard to understand the tolerance level of different overlapping modules and their molecular function of the protein-protein interaction networks. In this article, our focus on rigidity and flexibility analysis on protein-protein interaction subnetworks after detection of interference values from pair of key causative agents.  

\subsection*{Seed protein-protein interaction map and first order neighbors}

The seed network and first order network shows robustness and small world behavior. Lower clustering coefficient shows sparsity in interaction map as compared with cancer, aging subnetwork in human proteome \cite{ShaEtAl13}. Essential and hub proteins in seed network computed using various centrality measures such as linkedness of nodes, shortest path traversing through nodes, key connector proteins and maximally connected subgraphs demonstrates high number of genetic variants corresponding to the proteins {(see Supplementary Table S1)}. LMNA protein,a hub protein linked with 4 different neuro-muscular diseases. In our work, we found that LMNA protein receives interference from different pairs and also emits maximum interference to hub nodes at first order network when paired with other mutated proteins (see additional material 1). LMNA is suggested to play a role in nuclear stability, chromatin structure and gene expression\cite{CapCol06}. This infers towards the epigenetic regulation in progression of diseases by gene expression coordination in regulating different genes and subsequently to proteins in first order interactions as well as indirect interactions. Large number of proteins with high genetic variants shared different neuro-muscular diseases  which indicated towards co-morbidity in MDs.
Hub proteins with high genetic variants are positioned at central position and can possess lethal characteristics in diseased state of muscular dystrophies \cite{JeoEtAl01}. This lethality can disrupt the functionality of cells. Interconnected hub proteins of robust and small world first order MDs network displays high betweenness centrality values which indicates these proteins as maximum load bearing nodes in the network forming the core skeleton of the network. Core nodes of central module EEF1A1 constitutes of the same interconnected hubs, which suggests faster information flow at the core skeleton of network. Muscle interactome in \textit{Saccharomyces cerevisiae} and \textit{Caenorhabditis elegans} shows higher informational flow \cite{MisEtAl09}. Network propagation from essential and hub proteins from seed network demonstrates highly central proteins receiving mean interference and visits from different pair of complexes.

\subsection*{Structural integrity in protein-protein interaction networks}
Rigidity index in twelve subnetworks of the proteins that received large mean interference value (q$>90$)shows variability. The subnetworks constituting core module and core skeleton of first order networks shows highest rigidity. In details,our studies demonstrate the subnetwork of interconnected neighbors of EEF1A1 having hub property and core proteins of core module in first order network demonstrates the maximum rigidity. EEF1A1 Subnetwork demonstrated rigidity more than the parent network. Perturbing this subnetwork will largely affect the module EEF1A1 and other closely interconnected modules. This is because of the fact that we found  most of the proteins showing high bridgeness value, which gives account of inter-modular links constitutes this module. It is highly evident that this module is providing rigidity to the subnetwork of Human proteome related to muscular disorders.

In functional aspect, it is known that quality of muscles are affected by age and gender \cite{LynEtAl99} and the core node of this module is evidenced as core node of the core module of the Sirt family protein-protein interaction network, which is widely known to be implicated in aging related processes. However, Sirt7, a member of Sirt protein family, also showed mean interference and visits from the mutated complexes {(Supplementary Table S3)}. Sirt family of proteins also provides functionality related to muscle development \cite{ShaEtAl13}. Hormone replacement therapy is used for preserving muscle strength \cite{PhiEtAl10} and EEF1A1 module is associated with muscle degeneration with progression of age. Core node ESR1 (estrogen receptor alpha) of EEF1A1 module is involved in menopausal processes and therefore the muscle weakening in the women also emphasizes on this fact.

 The subnetworks of closely related 14.3.3 family protein YWHAE, YWHAG, YWHAZ, YWHAQ show rigidity and share large number of interacting partners, hence suggesting low tolerance level. Overall assessment of the rigidity in protein protein interaction network of the MDs evidenced towards core of the network being highly rigid with proteins having large number of visits from different mutated pairs. The perturbation in any member of 14.3.3. family subnetworks may influence connected modules and the other protein family members.
 
 Identification of genetic alterations that cause clinical phenotypes, such as severity of diseased states and mutations especially somatic are typically very diverse in neuro-muscular disorders. These are found in different subsets of genes in different patients\cite{ShoEtAl95}. The complexity grows with the heterogeneity of the mutations 
and their associations between individual mutations and a clinical phenotype. Driver mutations are observed to contribute to cancer development \cite{FutEtAl04}. Mutations driving neuro-muscular disorders may also target genes in cellular pathways and can influence first and second order interacting partners. Identification of driver nodes using network perspective \cite{YanEtAl11} will provide more knowledgeable insights on dynamics of tolerance level in different modules of networks.

\subsection*{Functional integrity} 
  
 Modular function related to core module of protein binding and protein domain specific binding is indispensable in MDs. Muscular dystrophies are largely known to be associated with inability in specific domain interactions in proteins \cite{StoEtAl05}. Network modular organization resulted in functionality related to protein binding and structure molecule activity, which emphasizes on the fact that network is functionally inclined towards the specific binding of cytoskeletal protein complexes to enhance structural integrity of the assembly of interacting partners.
Essential protein for protein synthesis EEF1A1 showed decline in protein abundance in wasted mutant mice leading  to  muscle wasting, motor neuron loss and immune system abnormalities \cite{KhaEtAl01}. The mutations in proteins can be considered as non functional entity in the network and thus considered as deleted which in a way can drive network towards homogeneity. Targeted or random deletion of nodes enhances homogeneity in network \cite{GhoEtAl13}. This homogeneity in network can proliferate the processes in unidirectional way. In other aspects of protein-protein interaction networks, modules become dysfunctional due to failure of key protein or simultaneously many proteins fail to perform function as in the case of mutations. Other modules or proteins have to share the burden of these modules or proteins or lack thereof. Hence it will be of interest to understand co-operativity issues related to driver mutations in network to which extent modules misses to perform functionality and which modules become performs functions to which extent. 

\section*{Conclusions}
In conclusion, PPI’s analysis not only reveals important characteristics and underlying behaviors, such as key interfered candidates but also tolerance level of networks and molecular mechanisms in muscular dystrophy. Hub nodes with large genetic variants involves in different neuro- muscular diseases suggesting comorbidity. Rigidity in networks is associated with the interconnected hubs in first order network. The core of the network faster informational flow with high betweenness value in interconnected hubs.
Functional rigidity in neuromuscular diseases is associated with protein binding and domain specific binding. This suggests muscular dystrophies may initiate due to failure of specific binding of the proteins which ultimately can affect interacting  proteins responsible for different molecular function such as cytoskeletal remodelling, protein folding and degradation, cell signalling modulation.

\begin{landscape}

\begin{table}[!ht]

   \renewcommand{\arraystretch}{1.5} 
   \caption{Function of the core nodes of the detected overlapping modules in first order network of seed mutated proteins}  
 \vspace{0.3cm}
  \label{tab:2}
        \begin{tabular}{|c|c|c|c|c|c|c|}
      
      \hline
     
     \textbf{Module} & \textbf{Module name}   & \textbf{Assignment}  & \textbf{GO-ID} & \textbf{p-value} & \textbf{Molecular function} & \textbf{Genes} \\  
        & & \textbf{value} & & & &  \\ 
  \hline

Module 1 & EEF1A1 & 534.90  & 5515 & 2.62E-02 & Protein binding  & EEF1A1, SUMO2, APP, YWHAZ, ,\\ 
       &&&&&& SUMO1, ESR1, ELAVL1, CUL1, FN1\\
Module 2 & CUL1 & 362.90  & 19899 & 1.81E-02 & Enzyme binding & CUL3, SUMO2, HSPA5, CUL1\\ 
Module 3 & ATM & 119.37  & 5198 & 2.78E-02 & Structural molecule & KRT9, KRT5, KRT10, TTN\\ 
&&&&&activity&\\
Module 4 & PRPF4 & 23.89  & 5515 & 4.04E-02 & Protein binding  & CUL3, APP, DDX23, ELAVL1, SNRNP40,\\ 
&&&&&& LSM2, PRPF4, CSTF1, TXNL4A, FN1\\
Module 5 & RBMBA & 8.24  & 3676 & 1.10E-02 &  Nucleic acid   & NUP153, APP, UPF3B, RBMBA, \\ 
&&&&&binding&ELAVL1, SFSA2, NFX1,SF2A3\\
Module 6 & SNRPF & 16.58  & 5515 & 3.31E-02 & Protein binding & CUL2, YWHAZ, SNRPD2,ELAVL1,\\ 
&&&&&&ITGA4,SNRPF,SNRPE, CUL1, FN1\\
Module7 & KRT1 & 91.18  & 5198 & 8.52E-06 & Strucural molecule & KRT9, KRT6A, KRT5, KRT14, \\ 
&&&&&activity& KRT1,KRT10\\ 
Module 8 & SMN1 & 10.90  & 32553 & 3.62E-02 & Ribonucleolide binding & PLK1, RAN, DDX20, HSPD1, HSPA9\\ 
Module 9 & HSPAB & 237.80  & 19904 & 4.01E-06 & Protein domain specific  & YWHAG, YWHAZ, YWHAH\\
&&&&& binding&HSP90AA1,YWHAQ, YWHAE\\
Module 10 & PCNA & 110.05  & 30528 & 2.10E-03 & Transcription regulator  & HDAC2, HDAC1, RAN\\ 
&&&&&activity&TP53, MYC, BRCA1\\ 
Module 11 & CSNK2A1 & 32.56  & 47485 & 2.50E-02 & Protein N-terminius  & CSNK2A1, PARP1\\
&&&&&binding&\\ 
 \hline
    \end{tabular}
    \end{table}
\end{landscape}

\begin{landscape}
\begin{table}[!ht]
   \centering
   
   \renewcommand{\arraystretch}{1.5} 
   \setcounter{table}{1} \renewcommand{\thetable}{\arabic{table}}
   \caption{  Continued :Function of the core nodes of the detected overlapping modules in first order network of seed mutated proteins} 
 \vspace{0.3cm}
 \label{tab:2}
        \begin{tabular}{|c|c|c|c|c|c|c|}
      
      \hline
     
     \textbf{Module} & \textbf{Module name}   & \textbf{Assignment}  & \textbf{GO-ID} & \textbf{p-value} & \textbf{Molecular Function} & \textbf{Genes} \\  
        & & \textbf{value} & & & &  \\  
  \hline

Module 12 & MEPCE & 21.66 & 47485 & 3.06E-02 & Protein N-terminius  & CSNK2A1, PARP1\\
&&&&&binding&\\ 
Module 13 & HSP90AA1 & 132.53  & 16740 & 3.77E-06 &Transferase activity  & EGFR, CDK1, MAPK1, PTK2, \\
&&&&&&PTK2B, PLK1, SRC, CDK2\\
Module 14 & RAC1 & 7.26  & 30234 & 3.87E-04 & Enzyme regulator  & GDI2, VAV2, ITSN1, ECT2, \\
&&&&&activity&KALRN \\
Module 15 & ATP5A1 & 34.17  & 5515 & 3.03E-02 & Protein binding & IKBKE, APP, SLC25A5, ATP5B,\\
&&&&&& CYCS, ATP5A1,SRC, CDK2, MDH2\\
Module 16 & NDUFA9 & 11.30  & 16491 & 5.69E-09 & Oxidoreductase  & NDUFB4, NDUFS6, NDUFA9, \\
&&&&&activity&NDUFB9, UQCRFS1, \\
&&&&&&NDUFA10M, UQCRB\\
Module 17 & NDUFS2 & 8.83  & 3824 & 8.29E-05 & Catalytic activity  & NDUFA5, NDUFB6, NDUFS4,\\
&&&&&& SUCLG2,  NDUFV2, NDUFS3,\\
&&&&&&  DLD, OGDH, NDUFS2, IDH3A\\
Module 18 & KRT85 & 4.76  & 5198 & 6.85E-04 & Structural molecule  & KRT81, KRT31, KRT85, KRT34, KRT33B\\
&&&&&activity&\\
Module 19 & FOXK1 & 2.18  & & & \textbf{ No Annotation}& \\
      \hline
    \end{tabular}
    \end{table}

  \end{landscape}

\begin{table}[!ht]
    \centering

     \renewcommand{\arraystretch}{1.5} 
\caption{The mean interference on the nodes originated from the duplets identified on the consensus basis on topological characteristics. The proteins in bold font show higher bridgeness value (greater than 1). \textit{q1} is the first quartile, \textit{q2} is the median and \textit{q3} is the third quartile.} 
   \vspace{0.6cm}
	\label{tab:3}
	\begin{tabular}{|cccc|} 
     
      \hline

        \textbf{Proteins with mean} &   \textbf{Proteins with mean}& \textbf{Proteins with mean} & \textbf{Proteins with mean}  \\ 
        \textbf{interference $\leq q1$ }&  \textbf{interference in $(q1, q2]$} & \textbf{interference in $(q2, q3]$}&  \textbf{interference $>q3$} \\ \hline
      ACTN1	&ACTA1	&ALB	&ACTA2\\
AHCYL1	& \textbf{ATP5B}	 & CBL	&ACTG1\\
ATM	& BRCA1	&CDK1&	ACTN2\\
BGN	&CAND1	&\textbf{CDK2}	&ACTN3\\
CAV1	&COPS5	& CTNNB1	& \textbf{ATP5A1}\\
CSK	&DNAJA1 &	F7	  &  CALM1\\
DDB1 &	EGFR & 	\textbf{FN1}	&CEP250\\
DRP2&	\textbf{HSP90AB1} &	HNRNPA2B1 &	\textbf{ESR1}\\
GCN1L1 &	\textbf{HSPB1}&	\textbf{HSP90AA1} &	H1F0\\
ITGB1	&\textbf{JUN}	&\textbf{HSPA8}	&HIST1H2AG\\
KRT6A	&KRT14 &	\textbf{HSPA9}	& HIST1H2BD\\
LAMA1	& KRT5&	\textbf{ITGA4}	&HNRNPC\\
LAMA5	& \textbf{MYC}	&\textbf{KIAA0101}	&HNRNPM\\
MAP2	& PRKCA &	KRT8 &	KRT18\\
NCSTN	& RUVBL2 &	MYL12A& MYH2\\
PRX	& SFN &	NONO	& MYH9\\
PSEN1	& SPTAN1	 & PXN	& MYL3\\
SGCA &	TPM3 &\textbf{	SIRT7} &	TSC2\\
SHC1	& \textbf{TUBB}	& TIAM1	& \textbf{VIM}\\
SNTA1	& UBR5 &	TPM1 &	YWHAE\\
TSC22D1	& VCAM1	& UBD	& YWHAH\\
VCL	&&	XRCC5	&\textbf{YWHAZ}\\
WWP1	&&&		\\
WWP2			&&&\\
\textbf{YWHAB}			&&& \\
       \hline

      \end{tabular}
      \end{table}

\clearpage


\section*{List of abbreviations used}
\begin{table}[!ht]

   \renewcommand{\arraystretch}{1.5} 
 \vspace{0.3cm}
      \begin{tabular}{cc}
      
      \hline
     
  Abbreviations &	Full name \\ 
  \hline
 
BC &	Betweenness centrality\\
CACNA1S &	Calcium channel, voltage-dependent, L type, alpha 1S subunit\\
CALM1	&Calmodulin\\
DMD	 	& Dystrophin\\
DNA		&Deoxyribonucleic acid\\
EEF1A1	&Eukaryotic Translation Elongation Factor 1 Alpha 1\\
ESR1		&Estrogen receptor alpha\\
GARS 	&	Glycyl-tRNA synthetase\\
GO-id	&	Gene Ontology ID\\
GYG1	&	Glycogenin 1\\
LGD, dominant &	Limb girdle dystrophies, dominant\\
LMNA	&	Lamin A/C\\
MCC		&Maximal clique centrality\\
MDs 		&	Muscular dystrophies\\
PPI's		&Protein-protein interaction\\
SGCA	&Sarcoglycan, alpha (35kDa dystrophin-associated glycoprotein\\
SGCG	&Sarcoglycan, gamma (35kDa dystrophin-associated glycoprotein\\
TTN		&Titin\\
YWHAE	&Tyrosine 3-monooxygenase/tryptophan 5-monooxygenase activation protein,\\   & epsilon polypeptide\\
YWHAG 	&Tyrosine 3-monooxygenase/tryptophan 5-monooxygenase activation protein, \\   & gamma polypeptide\\
YWHAQ	&Tyrosine 3-monooxygenase/tryptophan 5-monooxygenase activation protein, \\   & theta polypeptide\\
YWHAZ 	&Tyrosine 3-Monooxygenase/Tryptophan 5-Monooxygenase Activation Protein, \\   & zeta polypeptide\\

      \hline
\end{tabular}

\end{table}

\clearpage
\section*{Acknowledgement}
  \ifthenelse{\boolean{publ}}{\small}{}
Authors take pleasure in expressing gratitude to Dr. Naomi Fox of Lawrence Berkeley National Laboratory for her valuable discussion on the results on rigidity and flexibility issues on protein-protein interaction networks. The author also thanks his colleague, Kumar Parijat Tripathi for his valuable help in curation and conversion of data. This work has been partially funded by, the Italian Flagship Project {\em Interomics}, The Italian PON02\_00619 projects, and F.A.R. LAB-GTP initiative.

\section*{Competing Interest}
The authors declare that they have no competing interest.

\section*{Authors Contribution}
AS, MRG conceived of the study and designed the study with inputs from FDVB. AS, FM and MBF carried out the the studies. AS and MBF analyzed the results. AS, MBF and MRG drafted the manuscript. All authors read and approved the final manuscript. All authors read and approved the final manuscript.


\newpage
 \bibliographystyle{bmc_article}  
 \bibliography{bmc_article}      
%

\clearpage






%



\section*{Supplementary Figures}

\begin{figure}[!ht]

\setcounter{figure}{0} \renewcommand{\thefigure}{\arabic{figure}}
\renewcommand{\thefigure}{S\arabic{figure}}

	\centering
	 \includegraphics[scale=.3]{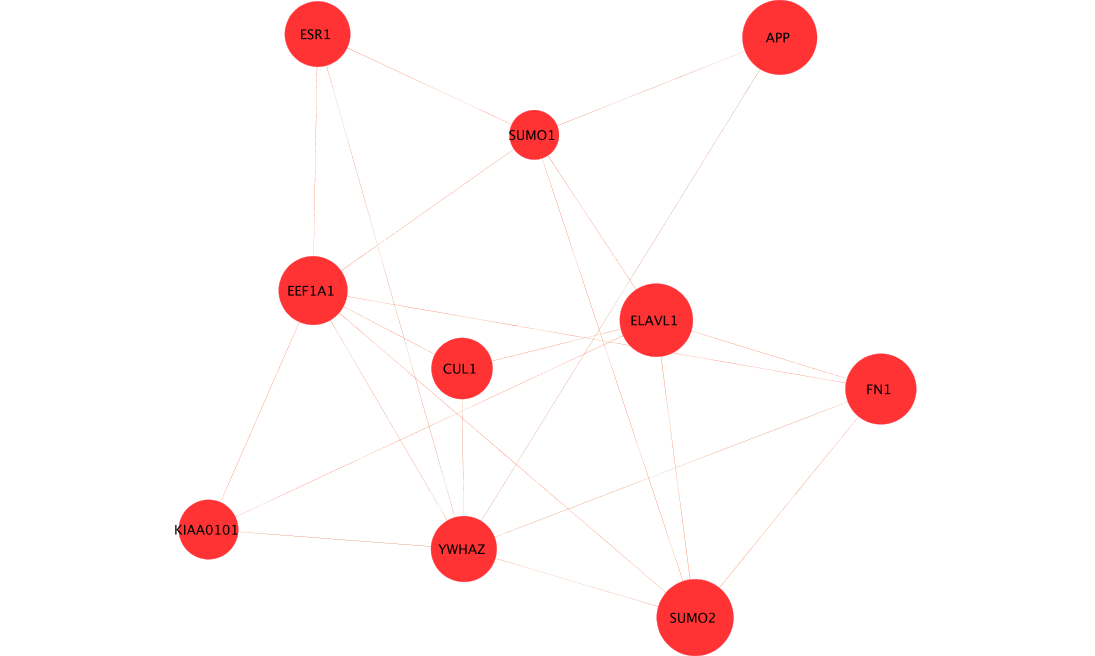}
\caption{Interconnected high degree proteins (hub nodes) in first order network.  Size of the node corresponds to the linkedness of the node.}       
 	\label{suppfig:1}
\end{figure}

\begin{figure}[!ht]
	
	\renewcommand{\thefigure}{S\arabic{figure}}
	\centering
	\includegraphics[scale=.6]{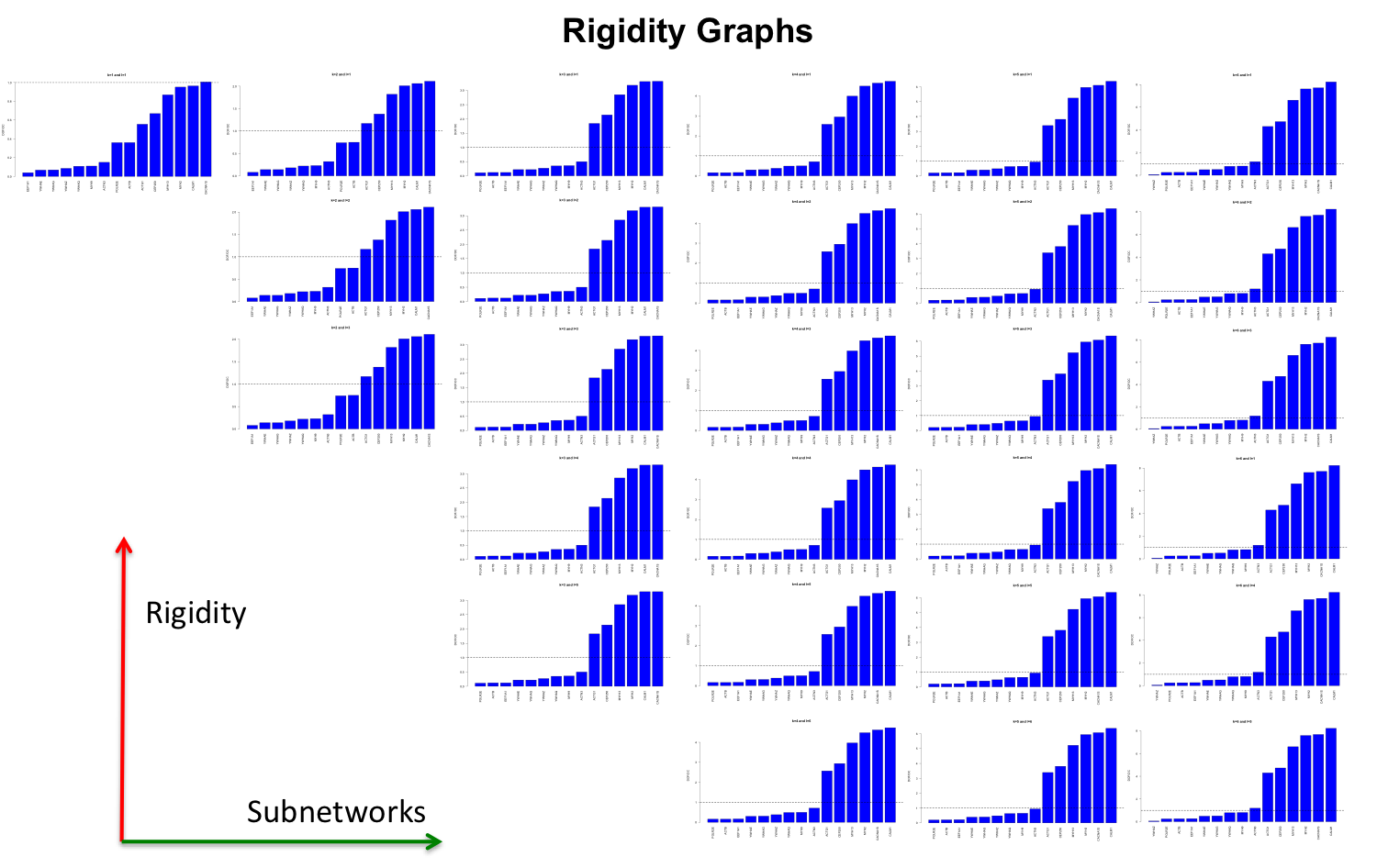}
\caption{Rigidity and flexibility analysis obtained from Kinari-LIB for all possible combination of $k=1$ and $l=1$ until $k=6$ and $l=6$ for  the subnetworks of nodes listed in table 2 which showed largest mean interference values calculated from emitting model of ITM Probe. }       
 	\label{suppfig:2}
\end{figure}  

\clearpage

\section*{Supplementary Tables}

	\begin{table}[!htbp]
	\renewcommand{\arraystretch}{1.3}

	\setcounter{table}{0} \renewcommand{\thetable}{\arabic{table}}
\renewcommand{\thetable}{S\arabic{table}}

   	\caption{Genetic variants associated causative genes of neuro-muscular diseases downloaded from Leiden Open Variant Database} 
   \vspace{0.6cm}
	\label{supptab:1a}

	\begin{tabular}{|cc||cc||cc|}
	\hline  
     
      Name & Number of variants & Name & Number of variants & Name & Number of variants\\
      \hline
ACTA1	& 378	&GAN	&6	&PLEC	&67\\
ACTC1	&142	&GARS	&107	&PLEKHG5&	0\\
AGRN	&32	&GDAP1	&0	&PMP22&	2\\
ANKRD1	&170&	GFPT1&	75	&POMGNT1&	244\\
ANO5	&336&	GJB1&	0&	POMT1&	396\\
ARHGEF10	&14&	GK&	176&	POMT2&	156\\
ASAH1	&19	&GMPPB&	27&	PRPS1&	0\\
ATL1	&13&	GNB4 &	11 &	PRX &	0\\
B3GALNT2&	28&	GNE&	741&	PTRF&	42\\
B3GNT1	&26&	GTDC2	&14	&RAB7A &	8\\
BAG3	&20	&HSPB1	&0	&RAPSN &	1008\\
BANF1	&38	&HSPB3	&0	&RYR1&	2116\\
BIN1	&70	&HSPB8&	0&	SBF2&	0\\
BSCL2	&0	&IGHMBP2&	172&	SEPN1&	1301\\
CAPN3	&2831 &	IKBKAP &	0	&SEPT9	 &0\\
CAV3	&467&	ISCU&	72&	SETX&	12\\
CCDC78&	5&	ISPD&	101&	SGCA&	753\\
CCT5	&13&	ITGA7&	179&	SGCB &	354\\
CFL2	 &44	&KBTBD13&	59 &	SGCD &	525\\
CHAT	&102	&KIF1B	&0	&SGCE	&323\\
CHKB	&37	&KLHL40&	56&	SGCG&	966\\
CHRNA1	&208&	LAMA2&	1503&	SGCZ&	16\\
CHRNB1	&198 &	LAMP2 &	97	&SH3TC2	&0\\
CHRND	&194 &	LARGE &	63	&SLC12A6 &	0\\
CHRNE	&418 &	LDB3	&101	&SMCHD1	&95\\
CNTN1	&8 &	LITAF &	0	&SMN1	&542\\
COL6A1	&313 &	LMNA	&3850	&SOX10	&0\\
COL6A2	&278	&MATR3&	123	&SPTLC1	&0\\
COL6A3	&364	&MFN2	&2&	SPTLC2	&0\\
COLQ	&211	&MICU1	&23	&SSPN	&16\\

       \hline

	\end{tabular}
 	\end{table}    
 
 \clearpage

\begin{table}[!htbp]
\renewcommand{\arraystretch}{1.3}
\centering
\setcounter{table}{0} \renewcommand{\thetable}{\arabic{table}}
\renewcommand{\thetable}{S\arabic{table}}
   	\caption{Genetic variants associated causative genes of neuro-muscular diseases downloaded from Leiden Open Variant Database. (Table continued)} 
   \vspace{0.6cm}
	\label{supptab:1b}

	\begin{tabular}{|cc||cc||cc|}
	\hline  
      
 Name & Number of variants & Name & Number of variants & Name & Number of variants\\
      \hline
      CRYAB	&95	&MPZ	&2	&SYNE1	&163\\
CTDP1	&0	&MSTN	&312	&SYNE2&	7\\
DAG1	&134	&MTM1&	529	&TCAP	&75\\
DCTN1&	0&	MTMR14	&36	&TMEM5&	36\\
DES	&305&	MTMR2	&0	&TNNI2&	61\\
DMD	&25828&	MUSK	&139	&TNNI3	&0\\
DMD\_d &	9235	&MYBPC3	&3	&TNNT1	&28\\
DNAJB6	&81	&MYH7	&4	&TNNT2	&1145\\
DNM2	&130 &	MYL2	&80	&TNNT3	&17\\
DOK7	&501	&MYL3&	74&	TNPO3&	65\\
DPM3	&5	&MYOT&	140	&TPM1	&177\\
DTNA	&1	&MYOZ1	&11	&TPM2	&517\\
DUX4	&159&	MYOZ2&	39	&TPM3	&78\\
DYSF	&2292	&MYOZ3	&19	&TRAPPC11	&8\\
EGR2	&0	&MYPN	&1474	&TRDN	&12\\
EMD	& 222	&NDRG1	&0	&TRIM32	&123\\
FAM134B	&0	&NEB	&243&	TTN	&3686\\
FGD4	&0	&NEFL	&3	&TTR	&0\\
FHL1	&70	&NGF	&0	&VCP	&161\\
FIG4	&0	&NTRK1&	724	&VMA21&	49\\
FKRP	&1007	&PABPN1&	520&	WNK1	&0\\
FKTN	&577	&PDK3	&26	&YARS	&0\\
FLNC	&39	&PDLIM3	&21	&ZMPSTE24&	1745\\
       \hline

	\end{tabular}
 	\end{table}  
 \clearpage
 
	\begin{table}[!htbp]
	\renewcommand{\arraystretch}{1.3} 
\setcounter{table}{1} \renewcommand{\thetable}{\arabic{table}}
\renewcommand{\thetable}{S\arabic{table}}
	\caption{Top core nodes of the overlapping modules detected by ModuLand framework}
	\label{supptab:2}
 \vspace{0.6cm}
      \begin{tabular}{|ccccccc|}
      \hline      
      \textbf{Module 1}	&\textbf{Module 2} &\textbf{	Module 3} &\textbf{	Module 4} &	\textbf{Module 5} &	\textbf{Module 6} &	\textbf{Module 7} \\
      \hline
      EEF1A1	& CUL3	&ATM	&PRPF4&	RBM8A&	SNRPF	&KRT1 \\ 
      SUMO2	&SUMO2 &	DDB1 &	LSM2 &	UPF3B &	SNRPE &	KRT14 \\
      ELAVL1	& COPS5	&TTN	&CSTF1 &	CDC40 &	SNRPD2	&KRT5 \\
      FN1	& CAND1 &	GCN1L1 &	SNRNP40 &	APP &	FN1 &	KRT6A \\
      ESR1	& CUL1&	DICER1&	TXNL4A &	NFX1 &	YWHAZ& 	KRT10 \\
      KIAA0101	& ELAVL1 &	DCD	& FN1	&FN1	&ELAVL1&	CDCP1 \\
      CUL1	&FN1 &	KRT10 &	APP &	ELAVL1 &	CDK2 &	KRT9\\
      APP	&APP &	RANBP2	&CUL3&	SF3A2&	ITGA4	&ATM \\
      YWHAZ	& ESR1&	KRT9&	ELAVL1& 	NUP153&	CUL2&	CBL\\
      SUMO1	&HSPA5 &	KRT5&	DDX23&	SF3A3&	CUL1	&GRB2\\
  
       \hline

        \hline
        
      \textbf{Module 8} & \textbf{Module 9} & \textbf{Module 10} & \textbf{Module 11} & \textbf{Module 12} & \textbf{Module 13} & \textbf{Module 14} \\
       \hline
       SMN1	&	HSPA8	&	PCNA	&	CSNK2A1	&	MEPCE	&	HSP90AA1	&	RAC1	\\
        DDX20	&	YWHAE	 &	S100A8	&	SART1	&	PRPF31	&	CDK1	&	GDI2	\\
        UBR5	&	HSP90AA1	&	HDAC1		&NUP188	&	CSNK2A1	&	PLK1		&RHOA	\\
        CALM1	&	YWHAQ	&	BRCA1	&	NUP93	&	PARP1	&	SRC	&	ECT2\\
        PLK1	&YWHAG	&	HDAC2		&PARP1	&	CSNK2B	&	EGFR	&	VAV2\\
        HSPA9	&	YWHAZ		&TP53	&	MEPCE	&	PLS3		&PTK2	&	ITSN1	\\
        RAN	&	YWHAH	&	RAN		&CSNK2B	&	PIN1	&	TUBB		&RHOD	\\
        PSMA3	&	CDK1	&	HSPB1	&	PRPF31	&	NUP93	&	CDK2	&	KALRN\\
        HSPD1	&	TUBB		&MDC1	&	PIN1	&	GFM1	&	MAPK1	&	RHOG\\
        MYC	&	PLK1		&MYC	&	PLS3		&MRPS16	&	PTK2B	&	RAC2\\
        \hline

        \hline
       
       \textbf{Module 15} & \textbf{Module 16} & \textbf{Module 17} & \textbf{Module 18} & \textbf{Module 19}& &\\
      \hline
      ATP5A1	&	NDUFA9	&	NDUFS2		&KRT85	&	FOXK1 & &\\
      ATP5B	&	NDUFA10	&	NDUFA5	&	KRT31	&	FOXK2 & &\\
      ATP5C1		&NDUFB4	&	IDH3A	&	KRT33B		&MPP7 & &\\
      APP	&	ATP5J2		&DLD	&	KRT34	&	SCLT1 & &\\
      MDH2	&	CYCS	&	OGDH	&	KRT81		&DYNLL2 & &\\
      IKBKE	&	UQCRH		&NDUFV2		&ADSL	&	LONP2 & &\\
      CDK2	&	NDUFB9	&	NDUFS3	&	USP15		&LIN7A & &\\
      SLC25A5	&	NDUFS6	&	NDUFB6	&	LGALS7&		NIN  & &\\
      SRC	&	UQCRFS1	&	NDUFS4	&	LRRC15&		BAG3 & &\\
      CYCS	&	UQCRB		&SUCLG2		&PGAM2	&	MPP5 & &\\
       \hline

\end{tabular}
\end{table}
     \clearpage 
      

\begin{table}[!htbp]

\renewcommand{\arraystretch}{1.3}
\renewcommand{\thetable}{S\arabic{table}}
\caption{Number of times proteins received interference and the mean inference of the node.}
	\label{supptab:3}
  \vspace{0.6cm}
	\begin{tabular}{|c|c|c||c|c|c|}
	\hline   
    Protein	&Number of duplets& 	Mean & Protein	& Number of duplets& 	Mean \\
   &that affect this protein&  interference 	& & that affect this protein& 	 interference\\
    \hline
YWHAZ	 &54	&0.03 &	ITGA4 & 9 &	0.02\\
ACTB & 50&	0.02	 &LARGE &9&	0.01\\
HSP90AA1 &	50&	0.02	&NCSTN&	9	&0.01\\
MYH9 &	47	&0.02	&PRX&	9	&0\\
YWHAG&	47&	0.02	&VCAM1&	9	&0.02\\
YWHAQ&	47&	0.02	&CAV3&	8	&0.01\\
YWHAE&	46	&0.02	&DRP2&	8	&0\\
APP&	44	&0.02	&MYH13&	8	&0.02\\
SUMO2&	44	&0.02	&MYL3&	8	&0.02\\
VIM&	44	&0.02	&EGFR&	7	&0.02\\
CALM1&	43	&0.02	&HSPD1&	7&	0.01\\
ELAVL1&	43	&0.02	&LAMA1&	7	&0\\
YWHAH&	43	&0.02	&LAMA5&	7&	0.01\\
ESR1&	42	&0.02	&DES&	6	&0.02\\
CBL&	41	&0.02	&HIST1H2BD&	6	&0.02\\
EEF1A1&	41	&0.02	&KRT1&	6	&0.02\\
ATP5A1&	39	&0.02	&MYH2&	6	&0.02\\
HNRNPC&	39	&0.02	&ACTN3&	5	&0.02\\
RPS3&	39	&0.02	&CDK1&	5	&0.02\\
ACTG1&	38&	0.02	&H1F0&	5	&0.02\\
FN1&	38	&0.02	&RALY&	5	&0.02\\
CUL3&	37	&0.02	&SDHB&	5	&0\\
TTN&	30	&0.02	&TPM3&	5	&0.02\\
HNRNPA2B1&	28	&0.02	&COPS5&	4	&0.02\\
MYL12A&	28&	0.02	&KRT10&	4	&0.02\\
H2AFX&	27	&0.02	&PXN&	4	&0.02\\
TUBB2A&	27	&0.02	&YWHAB&	4&	0.01\\
KRT8&	25 &	0.02	&ACTC1&	3	&0.02\\
CDK2&	22&	0.02	&ALB&	3	&0.02\\
HSPA5&	22&	0.02	&FLNA&	3	&0.02\\
KIAA0101&	22&	0.02	&HIST1H1D&	3&	0.02\\
KRT17&	22&	0.02	&HSPB1&	3	&0.01\\
GRB2&	21	&0.02	&KRT9&	3	&0.02\\
MYH11&	21	&0.02	&MAP2&	3&	0\\
ACTA2&	20	&0.02	&NONO&	3	&0.02\\
   \hline
\end{tabular}
\end{table}
       
   \clearpage   
      

    \begin{table}[!htbp]
    \renewcommand{\arraystretch}{1.3}
    \setcounter{table}{2} \renewcommand{\thetable}{\arabic{table}} 
    \renewcommand{\thetable}{S\arabic{table}}
    \caption{Number of times proteins received interference and the mean inference of the node. (Table continued)}
     \vspace{0.6cm}
    \label{supptab:3_continued}
    
      \begin{tabular}{|c|c|c||c|c|c|}
      \hline   
    Protein	&Number of duplets& 	Mean & Protein	& Number of duplets& 	Mean \\
   &that affect this protein&  interference 	& & that affect this protein& 	 interference\\
    \hline 
HSPA8&	20	&0.02	&ACTA1&	2	&0.02\\
TP53&	20&	0.02	&ACTN4&	2	&0.01\\
KRT18&	19&	0.02	&CAND1&	2	&0.02\\
CEP250&	18	&0.02	&DDB1&	2	&0.01\\
TSC2&	18	&0.02	&GAPDH&	2	&0.02\\
HIST1H1C&	17&	0.02	&GCN1L1&	2	&0.01\\
HSPA9&	17	&0.02	&HNRNPM&	2	&0.02\\
SRC&	17	&0.02	&POLR2E&	2	&0.02\\
TUBA1A&	17	&0.02	&S100A8&	2&	0.02\\
VCP&	17	&0.02	&SPTAN1&	2	&0.01\\
ACTN1&	15&	0.01	&TIAM1&	2	&0.02\\    
   CUL1&	15	&0.02	&UBD&	2	&0.02\\
LMNA&	15	&0.02	&XPO1&	2&	0.02\\
PTK2&	14	&0.01	&XRCC5&	2	&0.02\\
SUMO1&	13	&0.02	&AHCYL1&	1	&0.01\\
TUBB&	13	&0.02	&ATM&	1	&0.01\\
VCL&	13	&0.01	&ATP5B&	1&	0.02\\
ACTN2&	12	&0.02	&BRCA1&	1	&0.02\\ 
    CAV1&	12	&0.01	&CACNA1S&	1	&0.03\\
CSK&	12	&0.01	&CTNNB1&	1	&0.02\\
DMD&	12	&0.01	&DHX9&	1	&0.01\\
FYN&	12	&0.02	&DICER1&	1	&0.01\\
ITGB1&	12	&0.01	&DNAJA1&	1	&0.02\\
MYC&	12	&0.02	&F7&	1	&0.02\\
MYH7&	12	&0.02	&HSPA1L&	1	&0.02\\
NCK1&	12	&0.01	&JUN&	1	&0.01\\
PIK3R1&	12	&0.01	&KRT14&	1	&0.01\\
SHC1&	12	&0.01	&KRT5&	1	&0.01\\
SIRT7&	12	&0.02	&KRT6A&	1	&0.01\\
TLN1&	12	&0.01	&MAGI1&	1	&0\\
TPM1&	12	&0.02	&MYLK2&	1	&0.02\\
UTRN&	12	&0.01	&NCL&	1	&0.02\\
HIST1H2AG&	11	&0.02	&NCOA3&	1	&0.01\\

\hline
\end{tabular}
    \end{table}

\clearpage

\begin{table}[!htbp]
   \renewcommand{\arraystretch}{1.3}
    \setcounter{table}{2} \renewcommand{\thetable}{\arabic{table}} 
    \renewcommand{\thetable}{S\arabic{table}}
    \caption{Number of times proteins received interference and the mean inference of the node. (Table continued)}
     \vspace{0.6cm}
    \label{supptab:3_continued}
    
      \begin{tabular}{|c|c|c||c|c|c|}
      \hline   
    Protein	&Number of duplets& 	Mean & Protein	& Number of duplets& 	Mean \\
   &that affect this protein&  interference 	& & that affect this protein& 	 interference\\
    \hline 
    HSPG2&	11	&0.01	&PABPC1&	1	&0.01\\
PRKDC&	11	&0.02	&PRKCA&	1	&0.01\\
PSEN1&	11	&0.01	&RAC1&	1	&0.02\\
SGCA&	11	&0.01	&RUVBL2&	1	&0.02\\
SH3KBP1&	11&	0.01	&SFN&	1	&0.01\\
SNTA1&	11	&0.01	&TSC22D1&	1	&0.01\\    
VASP&	11	&0.02	&UBR5&	1	&0.02\\
BGN&	9	&0.01	&WWP1&	1	&0\\
HSP90AB1&	9	&0.02	&WWP2&	1	&0\\
    \hline
\end{tabular}
    \end{table}            
  \clearpage


\clearpage

\begin{table}[!htbp]
 \renewcommand{\arraystretch}{1.3}
\renewcommand{\thetable}{S\arabic{table}}
\caption{Proteins in first order network with maximum mean interference values from pairs of key proteins known as causative agents in muscular dystrophy.}     
   \label{supptab:4}
      \vspace{0.6cm}
     \begin{tabular}{|c|c|c|}
      \hline 
Proteins with mean interference $> q90$ & Number of duplets  affecting the protein & Mean interference\\
\hline
YWHAZ & 54 & 0.026\\
ACRB & 50 & 0.024\\
MYH9 & 47 & 0.022\\
YWHAG & 47 & 0.021\\
YWHAQ & 47 & 0.021\\
YWHAE & 46 & 0.021\\
CALM1 & 43 & 0.021\\
EEF1A1 & 41 & 0.02\\
ACTG1 & 38 & 0.02\\
CEP250 & 18 & 0.02\\
MYH13 & 8 & 0.021\\
MYH2 & 6 & 0.023\\
ACTN3 & 5 & 0.022\\
POLR2E & 2 & 0.02\\
CACNA1S & 1 & 0.025\\ \hline
\end{tabular}
\end{table} 
  
\clearpage
\begin{landscape}
\begin{table}
\renewcommand{\arraystretch}{1.3}
\renewcommand{\thetable}{S\arabic{table}}
\caption{DOF (Degrees  of freedom), overconstraints calculated using Pebble game algorithm for variable ($k=2$, $l=3$) and network topology of the subnetworks showing maximum mean interference from Emitting model of ITMPROBE for variable}
\label{supptab:5}
       \vspace{0.6cm}
      
       \centering
      \centering
      \begin{tabular}{|c|c|c|c|c|c|c|c|c|c|c|}
      \hline
      & DOF & Overconstraints & \% DOF vs & nodes & vertices & average degree & clustering & density & heterogeneity & path length\\
      &&& Overconstraints  &&&&coefficient &&&\\
      \hline
      EEF1A1 & 259400 & 21048 & 8.11 & 619 & 27173 & 87.79 & 0.60 & 0.14 & 0.65 & 1.85\\
      YWHAE & 11602 & 1624 & 13.99 & 406 & 12411 & 61.13 & 0.66 & 0.15 & 0.71 & 1.84\\
      YWHAG & 9205 & 1324 & 14.38 &  331 & 9864 & 59.60 & 0.65 & 0.18 & 0.69 & 1.81\\
      YWHAZ & 20186 & 3564 & 17.65 & 595 & 21371 & 71.95 & 0.55 & 0.12 & 0.76& 1.87\\
      YWHAQ & 10546 & 2371 & 22.48 & 395 & 11332 & 57.37 & 0.62 & 0.14 & 0.78 & 1.85\\
      MYH9 & 8060 & 1849 & 22.94 & 308 & 8672 & 56.31 & 0.62 & 0.18 & 0.60 & 1.81\\
      ACRN3 & 1160 & 367 & 31.63 & 61 & 1278 & 41.90 & 0.93 & 0.69 & 0.41 & 1.30\\
      POLR2E & 12162 & 8988 & 73.90 & 331 & 12801 & 79.75 & 0.71 & 24 & 0.569 & 1.75\\
      ACTB & 12715 & 9482 & 74.57 & 431 & 13574 & 62.98 & 0.53 & 0.14 & 0.75 & 1.85\\
      ACTG1 & 4760 & 5545 & 116.49 & 231 & 5218 & 45.17 & 0.62 & 0.19 & 0.698 & 1.63\\
      CEP250 & 5898 & 8140 & 138.01 & 185 & 6265 & 67.73 & 0.89 & 0.36 & 0.347 & 1.80\\
      MYH13 & 1389 & 2520 & 181.42 & 63 & 1512 & 48 & 0.92 & 0.77 & 0.26 & 1.22\\
      MYH2 & 1617 & 3242 & 200.49 & 90 & 1792 & 39.88 & 0.88 & 0.44 & 0.40 & 1.55\\
      CALM1 & 5091 & 10477 & 205.79 & 349 & 5779 & 33.11 & 0.55 & 0.09 & 0.902 & 1.90\\
      CACNA1S & 1130 & 2377 & 210.35 & 54 & 1234 & 45.70 & 0.96 & 0.86 & 0.268 & 1.13\\
       \hline

\end{tabular}
\end{table}
        
      \end{landscape}

%
%
%

\end{bmcformat}

\end{document}